\documentclass{aa}
\usepackage{graphicx}
\usepackage{amssymb}
\begin{document}

    \title{The large-scale anomalous microwave emission revisited by {\it WMAP} \thanks{The Wilkinson
Microwave Anisotropy Probe ({\it WMAP}) is the result of a partnership between Princeton
University and NASA's Goddard Space Flight Center}}

    \subtitle{}

    \author{G. Lagache\inst{1}
           }

    \offprints{Guilaine.Lagache@ias.u-psud.fr}

    \institute{IAS, B{\^{a}}t. 121, Universit\'e Paris-Sud, 91435 Orsay, France}

    \date{Received 24-03-03; Accepted 08-04-03}

\titlerunning{The ``anomalous'' microwave emission revisiting by {\it WMAP}}

\authorrunning{G. Lagache}

    \abstract{
We present a new study of the high latitude galactic contributions to the millimeter sky, based
on an analysis of the {\it WMAP} data combined with several templates of dust emission 
({\it DIRBE/COBE} and {\it FIRAS/COBE}) and gas tracers (HI and H$_{\rm \alpha}$).
To study the IR to millimeter properties of the diffuse 
sky at high galactic latitude, we concentrate on the emission correlated with the HI gas.
We compute the emission spectrum of the dust/free-free/synchrotron components associated with 
HI gas from low to large column densities.
A significant residual {\it WMAP} emission over the free-free, synchrotron and
the dust contributions is found from 3.2 to 9.1 mm.
We show that this residual {\it WMAP} emission (normalised to 10$^{\rm 20}$ atoms/cm$^{\rm 2}$)
(1) exhibits a constant 
spectrum from 3.2 to 9.1 mm and (2) significantly decreases in amplitude 
when N$_{\rm HI}$ increases,
contrary to the HI-normalised far-infrared emission which stays rather constant.
It is thus very likely that the residual {\it WMAP} emission is not
associated with the Large Grain dust component.
The decrease in amplitude with increasing opacity ressembles in fact to the decrease of the 
transiently heated dust grain emission observed in dense interstellar clouds.
This is supported by an observed decrease of the HI-normalised 60 $\rm \mu$m 
emission with HI column densities.
Although this result should be interpreted with care due to zodiacal residual contamination 
at 60 $\rm \mu$m, it suggests that the {\it WMAP} excess emission is associated with the small
transiently heated dust particles.
On the possible models of this so-called ``anomalous microwave emission'' linked 
to the small dust particles are the spinning dust and the excess
millimeter emission of the small grains, due to the cold temperatures
they can reach between two successive impacts with photons.
}
    \maketitle

\section{Introduction}\label{C}
At millimeter wavelengths, one of the major challenges in high 
sensitivity Cosmic Microwave Background (CMB)
anisotropy study is to determine the fraction of
the observed signal due to diffuse galactic foregrounds.
Three different components have been firmly identified
at high latitudes ($|b|>$10\degr):
thermal dust emission, synchrotron and free-free. 
Dust emission dominates the far-infrared surveys.
Its spatial distribution and frequency dependence
are quite well-determined for wavelengths shorter
than $\sim$800~$\rm \mu$m. Above $\sim$800~$\rm \mu$m, present data
currently do not give any strong constraints. 
So far, dust emission estimates in the millimeter range are
 thus an extrapolation
of what is known at shorter wavelengths.
Synchrotron radiation dominates radio-frequency surveys, but
Banday \& Wolfendale (\cite{Banday91}) and Bennett et al. (\cite{Bennett92})
showed that the spectral index steepens with frequency
and exhibits spatial variations which are poorly known.
Free-free emission has a well-determined spectral behavior
and templates are now available thanks to the
{\it WHAM} H$_{\rm \alpha}$ survey of the northern sky (Reynolds
et al. \cite{Reynolds98}, Haffner \cite{Haffner99}) and the {\it SHASSA} H$_{\rm \alpha}$ survey of the 
southern sky (Gaustad et al. \cite{Gaustad01}).\\

Cross-correlations of CMB data with far-infrared maps have revealed
the existence of a microwave emission component 
(the so-called ``anomalous microwave emission'') with
spatial distribution traced by these maps. This component
has a spectral index suggestive of free-free
emission and so has been first interpreted as free-free emission
(Kogut et al. \cite{Kogut96}). However, Kogut (\cite{Kogut99}) showed in small parts
of the sky that were covered by H$_{\rm \alpha}$ data 
that the microwave emission was consistently brighter than the free-free
emission traced by H$_{\rm \alpha}$.
Thus, the correlated component cannot be due to free-free emission alone. 
This is confirmed more recently by Banday et al. (\cite{Banday03}) also using {\it COBE/DMR}
data.\\

Recent works suggest that this anomalous far-infrared correlated component originates from spinning
dust grain emission (Draine \& Lazarian \cite{DL98a}, De Oliveira-Costa et al.
\cite{DOC99, DOC02}), tentatively detected at 5, 8 and 10 GHz by Finkbeiner et 
al. (\cite{Finkbeiner02}).
An alternative explanation is provided by thermal fluctuations in the
magnetization of interstellar grains causing magnetic dipole radiation 
(Draine \& Lazarian \cite{DL99}). However, very recently, Bennett et al. (\cite{Bennett03}) using 
{\it WMAP} data do not find any evidence for the anomalous microwave
emission. Their foreground component model comprises
only free-free, synchrotron and thermal dust emission,
and the observed galactic emission matches the model
to $<$1$\%$. Note that in their global analysis, they
are dominated by the brightest parts of the sky i.e.
the galactic plane and the high latitude dense interstellar
clouds. Thus, results may not apply to the most
diffuse regions. \\

We present in this paper a new study of the galactic
contributions to the millimeter sky, based
on an analysis of the {\it WMAP} data combined with several
templates of dust emission ({\it DIRBE/COBE} and {\it FIRAS/COBE}) 
and gas tracers (HI and H$_{\rm \alpha}$).
We focus only on the high latitude regions
where the results are easier to interpret in term
of physical properties of dust and where CMB
analysis are performed. 
The paper is organised as follows. We first present
the data we use together with their preparation
(Sect. \ref{data}). We then derive the spectrum
(from 100~$\rm \mu$m to 10 mm) of the HI-correlated
component (Sect 3.1) and show that there exists
a residual microwave emission (over free-free, synchrotron
and far-infrared dust emission)
whose HI-normalised amplitude decreases when the HI column density increases
but without any significant spectral variations 
(Sect. 3.2). We then discuss the results in Sect. 4.

\section{\label{data} Data-sets used and their preparation}

\subsection{{\it COBE} data}
The {\it COBE} satellite was developed by NASA's Goddard Space Flight Center 
to measure the diffuse infrared and microwave radiation from the early 
universe to the limits set by our astrophysical environment. 
It was launched November 18, 1989 and carried three instruments:
\begin{itemize}
\item a Far Infrared Absolute Spectrophotometer ({\it FIRAS}) to compare the
spectrum of the cosmic microwave background radiation with a 
precise blackbody (at each sky position, with an angular resolution
of 7\degr, we have one spectrum from 1 to 97 cm$^{\rm -1}$)
\item a Differential Microwave Radiometer 
({\it DMR}) to map the cosmic radiation (at 31, 53, 90 GHz
with a 7\degr beam)
\item a Diffuse 
Infrared Background Experiment ({\it DIRBE}) to search for
the cosmic infrared background radiation (10 photometric bands
from 1 to 240 $\rm \mu$m with an angular resolution of 40\arcmin). 
\end{itemize}

{\it COBE} data are presented in a quadrilateralized spherical projection
(the so-called {\it COBE} Quadrilateralized Spherical Cube, CSC), 
an approximately equal-area projection 
in which the celestial sphere is projected onto an inscribed cube. 
The {\it DIRBE} convention is to divide each cube face into 256$\times$256
pixels; thus all sky-maps have 256$\times$256$\times$6 = 393216 pixels.
Each pixel is approximatively 0.32\degr on a side.
For {\it FIRAS} and {\it DMR}, each cube face is 32$\times$32 pixels
leading to a total of 6144 pixels (of $\sim$2.6\degr).\\

We use the so-called (i) ``Sky Maps and Analyzed Science Data Sets'' {\it DMR} Data
(ii) ``Galactic Dust Continuum Spectra and Interstellar Dust Parameters'' {\it FIRAS} data, that 
give the residual sky spectrum after modelled emission from the CMB, zodiacal emission, 
and interstellar lines have been subtracted. 
(iii) {\it DIRBE} ``Zodi-Subtracted Mission Average (ZSMA) Maps'' for which
the zodiacal light intensities were subtracted week by week and the residual intensity 
values were averaged to create Maps. All {\it COBE} data are availabe
at http://lambda.gsfc.nasa.gov/product/cobe.\\

\subsection{Gas tracers}
The HI data we used are those of the Leiden/Dwingeloo survey which covers the
entire sky down to $\delta=-30^o$ with a grid spacing of 30' in both l and
b. The 36\arcmin\ half power beam width 
of the Dwingeloo 25m telescope provides 21 cm maps at an angular resolution
which closely matches that of the {\it DIRBE} maps. Details of the observations
and correction procedures are given by Hartmann (\cite{Hartmann94}) and by 
Hartmann $\&$ Burton (\cite{Hartmann97}). It should be noted that in this data-set
special care was taken for the removal of far sidelobes emission which makes
it particularly suitable for high latitude studies.
We derive the HI column densities
using 1~K~km~s$^{\rm -1}$=1.82~$10^{\rm 18}$~H~cm$^{\rm -2}$ (optically thin emission).\\

Thanks to the {\it WHAM} survey of the northern sky (Reynolds
et al. \cite{Reynolds98}, Haffner \cite{Haffner99}) and the {\it SHASSA} survey of the 
southern sky (Gaustad et al. \cite{Gaustad01}), it is now possible to have
a whole sky map of the H$_{\rm \alpha}$ emission (Dickinson et al. \cite{Dickinson03}, 
Finkbeiner \cite{Finkbeiner03}). Since the HI maps cover the sky down to
$\delta=-30\degr$, the H$_{\rm \alpha}$
emission we use is our analysis is mostly given by the {\it WHAM} survey.
{\it WHAM} provides a 12 km s$^{-1}$
velocity resolution with one-degree angular resolution
down to sensitivity limits of 0.2 R (1 R = 10$^{\rm 6}$$/$4$\pi$ ph cm$^{\rm -2}$ s$^{\rm -1}$ 
sr$^{-1}$) in a 30 second exposure.
The one-degree angular resolution nicely matches the 
{\it DIRBE} resolution.
We use the H$_{\rm \alpha}$ map and the conversion factors to free-free
emission (using T$_{\rm e}$=8000 K) from Finkbeiner (\cite{Finkbeiner03}) to derive templates
of free-free emission.
Since we work only on high latitude regions, the H$_{\rm \alpha}$
emission has not been corrected for extinction (the dust absorption
is likely to be very small, less than 5$\%$).
The free-free templates are used to derive a well-understood contribution
to the millimeter channels.

\subsection{Synchrotron templates}
Synchrotron emission arises from relativistic cosmic ray
electrons spiralling in the galactic magnetic field.
This emission dominates surveys at radio frequencies.
The only all-sky map at low frequencies that probe the 
synchrotron emission is the 408 MHz survey of Haslam et al.
(\cite{Haslam82}). For many years, this map has been used to predict
the synchrotron emission in the millimeter channels,
assuming a frequency dependence with a constant spectral index of about 2.75.
However, Bennett et al. (\cite{Bennett03}) have shown that the synchrotron spectral
index exhibit strong spatial variations and is steeper in the {\it WMAP}
bands than at radio frequencies. We thus use the {\it WMAP} synchrotron maps
derived by Bennett et al. (\cite{Bennett03}) using the Maximum Entropy Method 
as frequency dependent well-determined synchrotron templates.

\subsection{{\it WMAP} data}
The {\it WMAP}{\footnote{http://lambda.gsfc.nasa.gov/product/map/}} 
(Wilkinson Microwave Anisotropy Probe)
 mission, designed to determine the geometry, 
content, and evolution of the universe, has successfully provided
full sky maps at 23, 33, 41, 61 and 94 GHz at respectively
13.2, 39.6, 30.6, 21 and 13.2\arcmin\ FWHM resolution
with unprecedent sensitivity. A detailed description of the delivered
data-sets for the first 12 months of operation of {\it WMAP}
is given in the {\it WMAP: explanatory Supplement} (editor M. Limon
et al., Greenbelt, MD: NASA/GSFC).
The data we use are the first-year ``Smoothed I maps'' which are the temperature 
maps at each frequency, smoothed to a common resolution of 1 degree.
Data are delivered in the HEALPix{\footnote{http://www.eso.org/science/healpix}} 
format with N$_{\rm side}$=256

\subsection{\label{prep} Data preparation}
All the data have to be put in the same projection
and at the same resolution. The resolution is set by the {\it FIRAS}
experiment since it is the lowest resolution of our data-sets (7$ ^o$).
All data but {\it DMR} and {\it FIRAS} are thus converted in the {\it DIRBE} CSC format
and then convolved with the {\it FIRAS} beam and degraded
to the {\it FIRAS} CSC resolution (see Lagache \cite{Lagache03}
for more details). \\

We have removed for each data-set the cosecant
law variation (1) to avoid the obvious large scale correlations between
all galactic components concentrated in the disc 
and (2) to be consistent with the {\it WMAP}
data that measure only differentially on the sky and thus does
not measure the largest angular scales.\\

We restrict our analysis to $|b|>15\degr$ and exclude the Small
and Large Magellanic clouds, together with the $\rho$~-~Ophiucus
complex. We also remove cold molecular complexes (as the Taurus cloud),
and regions where the dust is locally
heated by nearby stars (like the HII regions)
following Lagache et al. (\cite{Lagache98}).
We stress out that this latter pixel selection,
although necessary to keep in the analysis
only diffuse parts of the sky,
does not change the results and conclusions of the paper.


\begin{figure*}
    \centering
    \includegraphics[width=15cm]{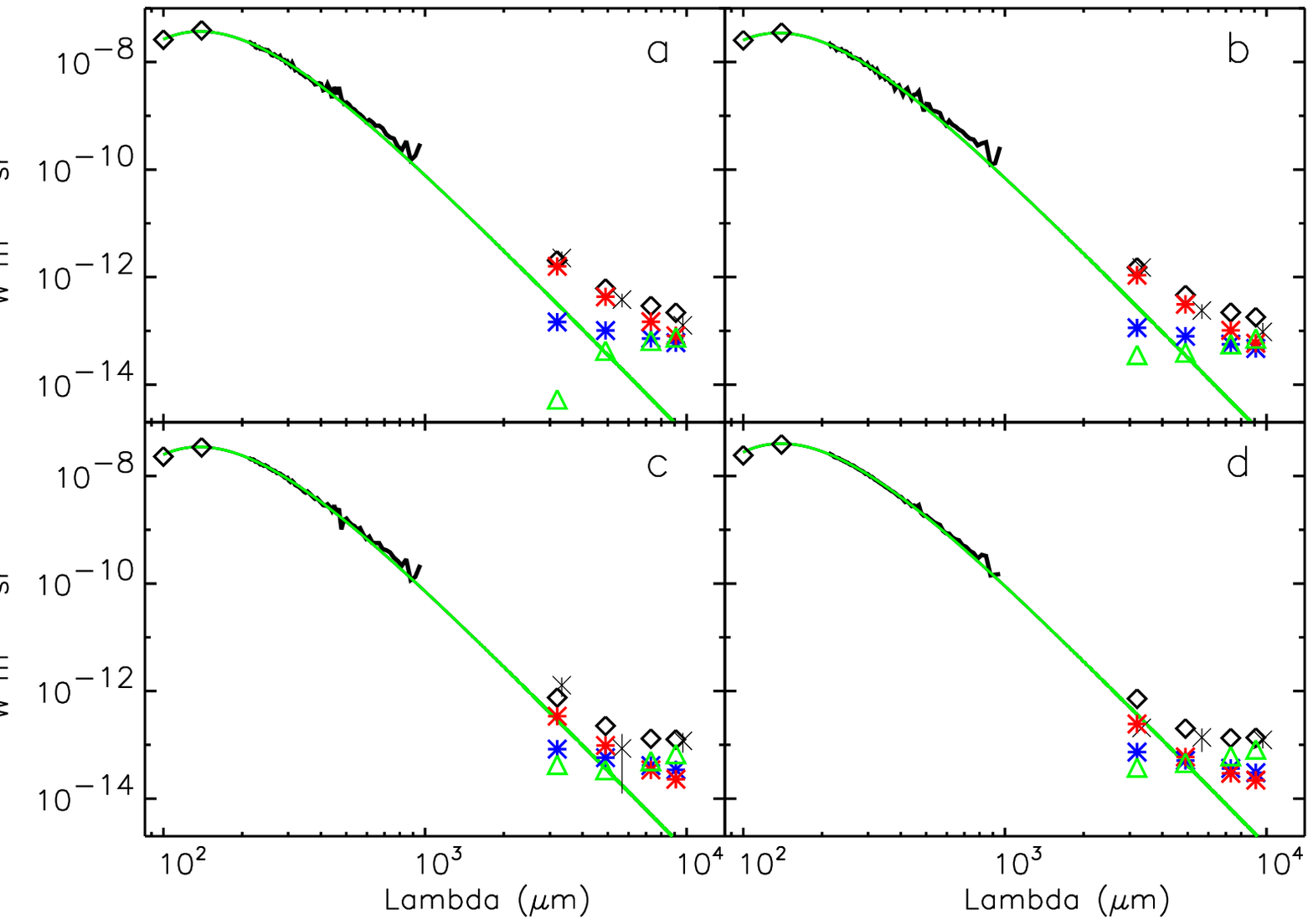}
    \caption{Spectrum of the HI-correlated component normalised to 10$^{\rm 20}$ at/cm$^{\rm 2}$ 
(with increasing N(HI) from a to d). The black diamonds at 100 and 140 $\rm \mu$m are the {\it DIRBE} data,
the black curve is the {\it FIRAS} spectrum, the black crosses with error bars
are the {\it DMR} data and the black triangles at 3.2, 4.9, 7.3 and 9.1 mm
are the {\it WMAP} data. Also displayed are the free-free and synchrotron
contributions (blue stars and green triangles respectively). The
green continous line is the result of a fit of the {\it DIRBE} 100, 140 $\rm \mu$m and {\it FIRAS} spectra 
($200<\lambda<500$ $\rm \mu$m) with a modified Planck curve with a $\nu^2$ emissivity law 
(the so-called stable thermal dust component). The residual
{\it WMAP} emission (which is the {\it WMAP}- free-free - synchrotron - stable thermal dust component) 
is shown as red stars. }
\label{spectra}
    \includegraphics[width=15cm]{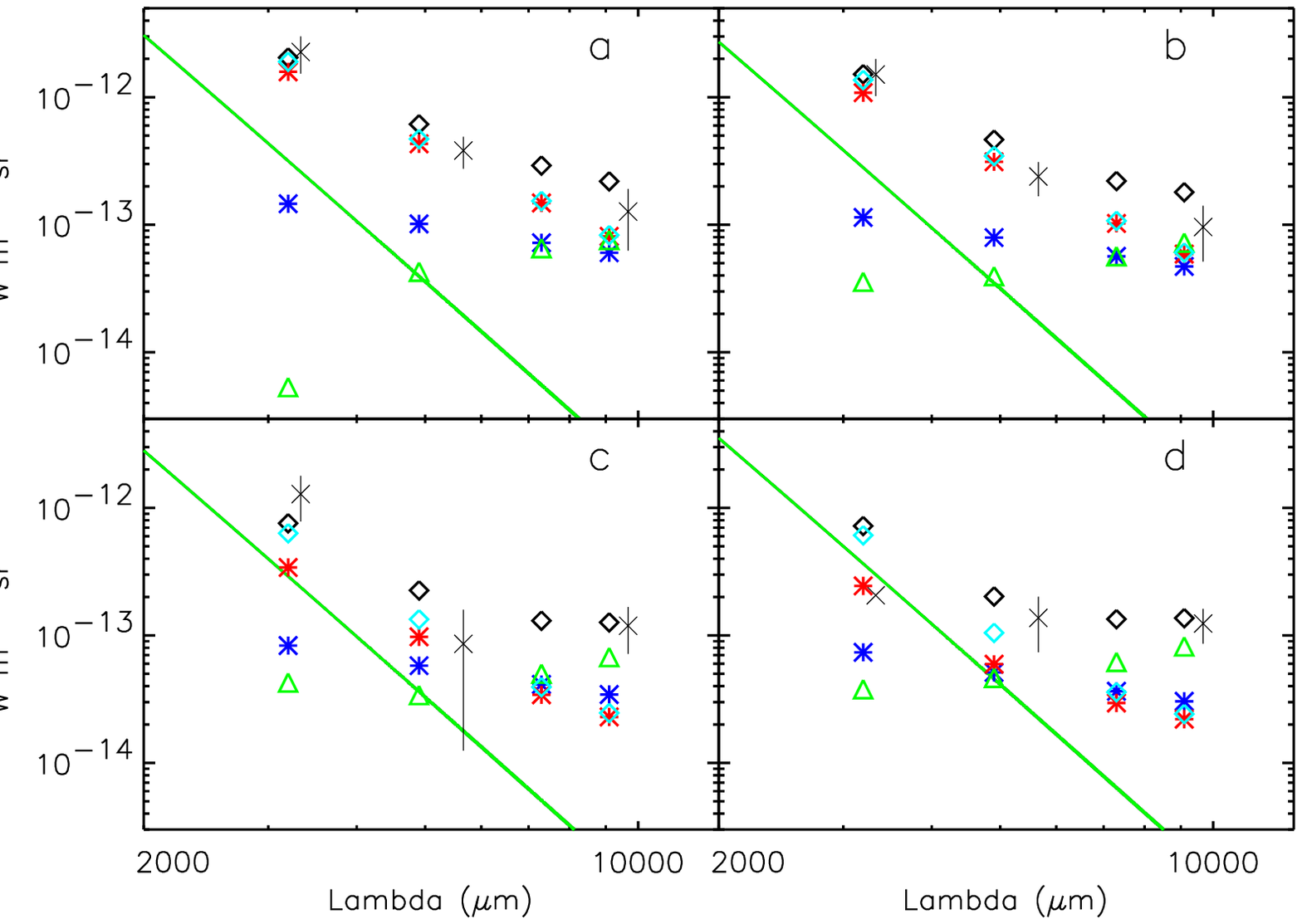}
    \caption{Zoom on the millimeter part of Fig. \ref{spectra}. Symbols and colors
are the same as on Fig. \ref{spectra}. Added is the residual {\it WMAP} emission
after having removed only the free-free and synchrotron contributions (light-blue diamonds).}
\label{zoom_spectra}
\end{figure*}

\begin{table*}
\begin{center}
\begin{tabular}{|ll|l|l|l|l|} \hline
Component  &  & 3.2 mm &  4.3 mm & 7.9 mm & 9.1 mm \\ \hline
{\it  WMAP} & 1 & 2.05 10$^{\rm -12}$ & 6.15 10$^{\rm -13}$ & 2.92 10$^{\rm -13}$ & 2.19 10$^{\rm -13}$ \\
          &  2 & 1.52 10$^{\rm -12}$ & 4.65 10$^{\rm -13}$ & 2.25 10$^{\rm -13}$ & 1.80 10$^{\rm -13}$ \\
          &  3 & 7.59 10$^{\rm -13}$ & 2.26 10$^{\rm -13}$ & 1.31 10$^{\rm -13}$ & 1.27 10$^{\rm -13}$ \\
          &  4 & 7.22 10$^{\rm -13}$ & 2.03 10$^{\rm -13}$ & 1.34 10$^{\rm -13}$ & 1.37 10$^{\rm -13}$ \\ \hline

{\it Free-free} &  1 & 1.46 10$^{\rm -13}$ &  1.02 10$^{\rm -13}$ & 7.23 10$^{\rm -14}$ & 6.03 10$^{\rm -14}$  \\
                &  2 & 1.14 10$^{\rm -13}$ &  7.93 10$^{\rm -14}$ & 5.65 10$^{\rm -14}$ & 4.71 10$^{\rm -14}$ \\
                &  3 & 8.33 10$^{\rm -14}$ &  5.79 10$^{\rm -14}$ & 4.12 10$^{\rm -14}$ & 3.44 10$^{\rm -14}$ \\
                &  4 & 7.36 10$^{\rm -14}$ &  5.12 10$^{\rm -14}$ & 3.64 10$^{\rm -14}$ & 3.04 10$^{\rm -14}$ \\ \hline

{\it Synchrotron}& 1 & 5.32 10$^{\rm -15}$ & 4.28 10$^{\rm -14}$ & 6.56 10$^{\rm -14}$ & 7.60 10$^{\rm -14}$ \\
                 & 2 & 3.57 10$^{\rm -14}$ & 3.95 10$^{\rm -14}$ & 5.67 10$^{\rm -14}$ & 7.24 10$^{\rm -14}$ \\
                 & 3 & 4.27 10$^{\rm -14}$ & 3.40 10$^{\rm -14}$ & 5.01 10$^{\rm -14}$ & 6.76 10$^{\rm -14}$ \\
                 & 4 & 3.79 10$^{\rm -14}$ & 4.64 10$^{\rm -14}$ & 6.19 10$^{\rm -14}$ & 8.20 10$^{\rm -14}$ \\ \hline

{\it Stable thermal} & 1 & 3.17 10$^{\rm -13}$ & 3.95 10$^{\rm -14}$ & 5.53 10$^{\rm -15}$ & 1.86 10$^{\rm -15}$ \\ 
{\it dust}           & 2 & 2.81 10$^{\rm -13}$ & 3.50 10$^{\rm -14}$ & 4.90 10$^{\rm -15}$ & 1.64 10$^{\rm -15}$ \\
                     & 3 & 2.91 10$^{\rm -13}$ & 3.62 10$^{\rm -14}$ & 5.08 10$^{\rm -15}$ & 1.71 10$^{\rm -15}$ \\
                     & 4 & 3.66 10$^{\rm -13}$ & 4.55 10$^{\rm -14}$ & 6.39 10$^{\rm -15}$ & 2.15 10$^{\rm -15}$ \\ \hline

{\it Residue}    & 1 & 1.58 10$^{\rm -12}$ & 4.31 10$^{\rm -13}$ & 1.48 10$^{\rm -13}$ & 8.11 10$^{\rm -14}$ \\
                 & 2 & 1.09 10$^{\rm -12}$ & 3.12 10$^{\rm -13}$ & 1.02 10$^{\rm -13}$ & 5.89 10$^{\rm -14}$ \\
                 & 3 & 3.41 10$^{\rm -13}$ & 9.75 10$^{\rm -14}$ & 3.42 10$^{\rm -14}$ & 2.29 10$^{\rm -14}$ \\
                 & 4 & 2.45 10$^{\rm -13}$ & 5.94 10$^{\rm -14}$ & 2.95 10$^{\rm -14}$ & 2.20 10$^{\rm -14}$ \\ \hline

\end{tabular}
\end{center}
\caption{\label{table_spectra} {\it WMAP}, free-free, synchrotron and stable thermal
dust component emission (in W/m$^{\rm 2}$/sr, normalised to 10$^{\rm 20}$ at/cm$^{\rm 2}$) in the four 
HI bins together with the residual emission (which is equal
to  {\it WMAP} - Free-free - Synchrotron - Stable thermal dust). The mean total HI column densities
are 3.3, 4.1, 5.6 and 9.9 10$^{\rm 20}$ at/cm$^{\rm 2}$ for the bin 1, 2, 3 and 4 respectively.}
\end{table*}

\section{Analysing the data}

\subsection{Deriving the HI-correlated component spectrum}
The HI-correlated dust emission is the dominant component
at high galactic latitude at infrared/far-infrared/submillimeter 
wavelengths
(except in the very low HI column density regions where the 
Cosmic Infrared Background
becomes an important contribution, e.g. Lagache et al. \cite{Lagache99}).
One way to study the infrared to millimeter properties of the diffuse 
sky at high galactic latitude is therefore to concentrate on 
the emission correlated with the HI gas. 
We search here for the spatial/spectral variations of the infrared to
millimeter emission with the HI gas column densities.\\

To compute the emission spectrum of the component associated with 
HI gas from low to large column densities, we use a 
differential method that removes, within statistical variance, any residual infrared emission 
that is not correlated with the HI gas such as an
isotropic component.
We first select sky pixels according to their
HI column density and sort them into sets of pixels bracketed by selected values of
$N_{HI}$.  
Correlated HI emission spectra are then computed for each set of pixel k using the equation:
\begin{equation}
F_{\nu}(k) = \frac{<F>_k - <F>_0}{<N_{HI}>_k - <N_{HI}>_0}
\label{eq_spec}
\end{equation}
where $<F>_{i}$ corresponds to the mean emission computed for
the set of pixels i, and  $<N_{HI}>_{i}$ to the mean HI column density for
the same set of pixels. Note that all the data-sets used here are cosecant-law
subtracted (see Sect. \ref{prep}).\\

To keep high signal-to-noise ratio, only 5 sets of pixels are considered here, 
with increasing $N_{HI}$. The first set (labeled ``0'' in Eq. (\ref{eq_spec})), 
serves as the ``reference'' set and corresponds to the lowest column 
density regions (representing $\sim5\%$ of the sky).
We are thus left with 4 sets of pixels k
with increasing $N_{HI}$ and derive accordingly four mean spectra $F_{\nu}(k)$.
The sets of pixels are selected on the cosecant-law removed HI emission
that can be negative. For reference, the total mean HI column
density (i.e. non cosecant-law subtracted) 
for the 4 bins are 3.3, 4.1, 5.6 and 9.9 10$^{\rm 20}$ at/cm$^{\rm 2}$.  
By construction, the spectra are normalised to 
10$^{\rm 20}$ at/cm$^{\rm 2}$. 
Note that F in  Eq. (\ref{eq_spec}) represents alternatively the {\it DIRBE}, {\it FIRAS}, {\it DMR}, {\it WMAP}, free-free 
and synchrotron data. \\

\subsection{Results}
The four spectra are presented on Fig. \ref{spectra} with the {\it DIRBE} data points
at 100 and 140~$\rm \mu$m, the {\it FIRAS} spectra (displayed only between 210 and
1000 $\rm \mu$m), the {\it DMR} data points at 90, 53 and 31 GHz and the {\it WMAP}
data points at 3.2, 4.9, 7.3 and 9.1~mm (all these data points are in black
on Fig. 1). A zoom on the millimeter part of the figure is presented
on Fig. \ref{zoom_spectra}.
We fit the {\it DIRBE} 100, 140~$\rm \mu$m and {\it FIRAS} spectra ($200<\lambda<500$~$\rm \mu$m) 
with a modified Planck curve with a $\nu^2$ emissivity law (the result
of the fit is displayed on Fig. \ref{spectra} and \ref{zoom_spectra}). We know that this fit
is inconsistent with {\it FIRAS} data below $\sim$500~GHz where
an excess component is detected (Reach et al. \cite{Reach95}, Finkbeiner
et al. \cite{Finkbeiner99}). However, discussing this component is not the goal
of this paper. 
We only concentrate on the millimeter part on the spectra and how it
relates to the far-infrared emission. It is important to note that this
far-infrared dust emission has a stable spectrum, not changing with increasing
opacity (Lagache et al. \cite{Lagache99}).
In this framework, the 
$\nu^2$ modified black body is well representative
and useful for comparison between spectral far-infrared and millimeter 
shapes{\footnote{The way we are fiting the far-infrared stable component
is not critical since we focus on the variable millimeter emission.}}. 
This far-infrared dust emission extrapolated at millimeter wavelengths
will be called the ``stable thermal dust component''.\\

First, we see on Fig. \ref{zoom_spectra} that there is a strong millimeter 
excess (with both {\it DMR} and {\it WMAP} data) 
with respect to the stable thermal dust component (i.e the $\nu^2$ modified black body).
This excess decreases significantly (by a factor of about 5 at 3.2 mm)
when the HI column density increases,
although the far-infrared emission remains nearly constant (at the $\sim$6$\%$ level).
The far-infrared emission is dominated by the so-called Large Grain dust
component. The millimeter excess, which changes rapidely with opacity, is thus not 
likely associated with this dust component.\\

\begin{table*}
\label{Table_2}
\begin{center}
\begin{tabular}{l|c|c|c|c} 
  & HI bin 1 & HI bin 2 & HI bin 3 & HI bin 4 \\ \hline \hline
$\rm \nu R_{\nu}$(3.2 mm)  &  1.58 10$^{\rm -12}$ & 1.09 10$^{\rm -12}$ & 3.41 10$^{\rm -13}$ & 2.45 10$^{\rm -13}$ \\ \hline
$\rm \nu R_{\nu}$(4.9 mm)  &  4.31 10$^{\rm -13}$ & 3.12 10$^{\rm -13}$ & 9.75 10$^{\rm -14}$ & 5.94 10$^{\rm -14}$  \\ \hline
$\rm \nu R_{\nu}$(7.3 mm)  &  1.48 10$^{\rm -13}$ & 1.02 10$^{\rm -13}$ & 3.42 10$^{\rm -14}$ & 2.95 10$^{\rm -14}$  \\ \hline
$\rm \nu R_{\nu}$(9.1 mm)  &  8.11 10$^{\rm -14}$ & 5.89 10$^{\rm -14}$ & 2.29 10$^{\rm -14}$ & 2.20 10$^{\rm -14}$  \\ \hline \hline
$\rm \nu I_{\nu}$(240 $\rm \mu$m) &  1.81 10$^{\rm -8}$ & 1.63 10$^{\rm -8}$ & 1.59 10$^{\rm -8}$  & 1.83 10$^{\rm -8}$  \\ \hline
$\rm \nu I_{\nu}$(140 $\rm \mu$m) &  3.90 10$^{\rm -8}$ & 3.54 10$^{\rm -8}$ & 3.41 10$^{\rm -8}$  & 3.90 10$^{\rm -8}$  \\ \hline
$\rm \nu I_{\nu}$(100 $\rm \mu$m) &  2.61 10$^{\rm -8}$ & 2.54 10$^{\rm -8}$ & 2.31 10$^{\rm -8}$  & 2.43 10$^{\rm -8}$ \\ \hline
$\rm \nu I_{\nu}$(60 $\rm \mu$m) $|\beta|>3\degr$           & 1.16 10$^{\rm -8}$ & 9.37 10$^{\rm -9}$ & 7.18 10$^{\rm -9}$  & 6.48 10$^{\rm -9}$  \\
$\rm \nu I_{\nu}$$_{\rm VSG}$(60 $\rm \mu$m) $|\beta|>3\degr$   & 9.24 10$^{\rm -9}$ & 6.98 10$^{\rm -9}$ & 4.94 10$^{\rm -9}$  & 4.21 10$^{\rm -9}$  \\
$\rm \nu I_{\nu}$(60 $\rm \mu$m) $|\beta|>15\degr$          & 9.34 10$^{\rm -9}$ & 7.23 10$^{\rm -9}$ & 4.57 10$^{\rm -9}$  & 4.92 10$^{\rm -9}$  \\
$\rm \nu I_{\nu}$$_{\rm VSG}$(60 $\rm \mu$m)  $|\beta|>15\degr$ & 7.01 10$^{\rm -9}$ & 4.83 10$^{\rm -9}$ & 2.32 10$^{\rm -9}$  & 2.65 10$^{\rm -9}$  \\ 
\end{tabular}
\end{center}
\caption{Excess residual emission ($\rm \nu R_{\nu}$) at 3.2, 4.9, 7.3 and 9.1 mm
with 240, 140, 100 associated brightnesses (in W/m$^{\rm 2}$/sr, normalised to 10$^{\rm 20}$ at/cm$^{\rm 2}$) 
for the 4 HI bins. The 60 $\rm \mu$m brightness is given
for $|\beta|>3\degr$ and  $|\beta|>15\degr$ to show that the decrease may not be
due to any residual zodical light emission. Also given are the 60 $\rm \mu$m
brightnesses corresponding to the Very Small Grains dust component only
(we have removed from the 60 $\rm \mu$m emission, I(60 $\rm \mu$m), the best $\nu^2$ modified
black body fit done on the Large Grain dust component).}
\end{table*}

We can go further by removing to the {\it WMAP} emission the corresponding
free-free, synchrotron and stable thermal dust component contribution.
The residual {\it WMAP} emission
is shown on Fig. \ref{spectra} and \ref{zoom_spectra} (red stars) and detailed in Table \ref{table_spectra}.
First, at each frequency, the residual emission exhibits a strong decrease (by about a factor
of 5) with HI column densities (from bin 1 to 4). Second, the residual emission
decreases from 3.2 to 9.1 mm in each HI bin.  
On Fig. \ref{resi_spectra} are shown
the {\it WMAP} residual emissions for the 4 bins at 3.2, 4.9, 7.3 and 9.1 mm, 
normalised to the 90 GHz {\it DMR} residual emission (the 31 and 53 GHz {\it DMR} residual emissions
have also been computed but are not displayed to avoid confusion. Results, although
more noisy, are in very good agreement with {\it WMAP}). This figure shows that
we do not detect any significant variations in the spectral shape
of the residual emission{\footnote{This however will have to be quantified
when smoothed {\it WMAP} data with error bars will be available.}.
Thus, the HI-normalised residual emission, although decreasing in amplitude with the
HI column density, has a constant spectrum.

\section{Discussion}
To account for the galactic energy emitted from the mid-infrared to the submillimeter, it
is necessary to have a broad dust size distribution from large grains
down to large molecules. For example, D\'esert et al. (1990) (see also Draine \& Anderson \cite{Draine85}, 
Puget et al. \cite{Puget85}, Weiland et al. \cite{Weiland86}, Siebenmorgen \& Kr\"ugel \cite{Sieben92}, 
Dwek et al. \cite{Dwek97} and more recently Li \& Draine \cite{Li01}) have proposed a consistent interpretation of both
the infrared emission in diffuse HI clouds and the interstellar extinction
curve using a model with three components: PAHs (Policyclic Aromatic
Hydrocarbons), Very Small Grains (VSGs) and Large Grains. 
PAHs and VSGs are small enough ($a\le10$ nm) to experience significant temperature fluctuations
after photon absorption. They emit over a wide range of temperatures and dominate
the emission for $\lambda \le 60$~$\rm \mu$m.
The Large Grain component is the more traditional dust component historically inferred
from optical studies. These grains are in
equilibrium with the incident radiation field with a temperature
of about 17~K in the diffuse atomic medium (Boulanger et al. \cite{Boulanger96}).
The Large Grain dust component is expected, at long wavelengths, to
be proportional to the total amount of solid material.
The large spatial variations of the infrared spectrum over the 
wavelenghts range 12-60~$\rm \mu$m have been interpreted as changes 
in the abundance of small grains (Boulanger et al. \cite{Boulanger90}, 
Laureijs et al. \cite{Laureijs91},
Bernard et al. \cite{Bernard93}, Abergel et al. \cite{Abergel94}). In particular, 
strong deficits of the transiently heated grains emitted at 60~$\rm \mu$m
are observed in dense interstellar clouds, these deficits being explained
by grain-grain coagulation processes (e.g. Stepnik et al. \cite{Stepnik03}).\\

\begin{figure}
    \includegraphics[width=9.5cm]{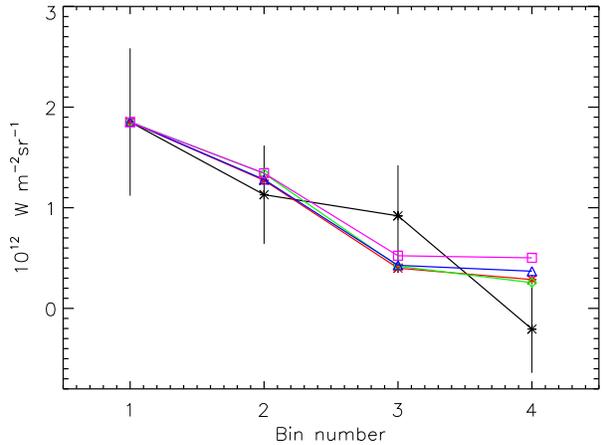}
    \caption{HI-normalised residual {\it WMAP} emission at 3.2 (red), 4.9 (green), 7.3 (blue)
and 9.1 (magenta) mm. All spectra are normalised on the {\it DMR} 90 GHz
residual emission for the first HI bin (black points with error bars).
The HI column density increases from bin 1 to 4 (from 3.3 to 9.9 10$^{\rm 20}$ at/cm$^{\rm 2}$). 
No significant variations 
in the spectral shape of the residual emission is detected.}
\label{resi_spectra}
\end{figure}

It has been shown in Sect 3.2 that the HI-normalised residual
{\it WMAP} emission (i.e. the excess above free-free
and synchrotron contributions and the stable thermal dust component) 
is well traced at large
scale by the HI gas and (1) exhibits a constant 
spectrum from 3.2 to 9.1 mm but (2) significantly decreases in amplitude 
when N$_{\rm HI}$ increases,
contrary to the far-infrared emission (associated with 
the so-called stable thermal dust component)
which stays rather constant (cf. Table 2).
It is thus very likely that the residual {\it WMAP} emission is not
associated with the Large Grain dust component.
The decrease in amplitude ressembles in fact to the decrease of the 
PAH/VSGs emission observed in dense interstellar clouds. By extrapolating
the PAH/VSGs behaviour from dense interstellar clouds to the 
diffuse medium, we may expect, when increasing the HI
column density, to decrease the PAH and VSGs proportion
and thus the mid-infrared emission.
If this is true, then the PAH/VSGs HI-correlated
emission should decrease with HI column densities.
This decrease, if present, is very hard to observe
in the mid-infrared due to the strong residual interplanetary
dust emission at large scale. On the {\it DIRBE} bands,
only the 60 $\rm \mu$m may be used. We have computed for the 4
HI bins the 60 $\rm \mu$m HI-correlated emission
with two different cuts in ecliptic latitude ($|\beta|>3\degr$
and $|\beta|>15\degr$). Although the absolute level of the 
60 $\rm \mu$m HI-correlated emission varies for the 2 cuts,
we observe nearly the same significant decrease of the 60 
$\rm \mu$m emission with the HI column density (cf. Table 2).
The 60 $\rm \mu$m band may be contaminated by the Large
Grain emission (30 to 40 $\%$, e.g. D\'esert et al. 1990).
Therefore, we remove to the 60 $\rm \mu$m emission the Large Grain
contamination using the best $\nu^2$ modified
black body fit (Fig. 1). The observed decrease
at 60 $\rm \mu$m becomes even larger (Table 2).
Although this result should be interpreted with
care due to the zodiacal contamination
at 60 $\rm \mu$m, it suggests that the {\it WMAP}
residual emission is associated with the small
transiently heated particles.\\

The previous results suggest the anomalous
microwave component is associated
with the transiently heated dust particles,
but its exact physical mechanism remains to be
found. On the possible models of the anomalous
emission linked to the transiently heated particles are:
\begin{itemize}
\item The ``spinning-dust'' which is the rotational emission
from very small dust grains (Draine \& Lazarian, \cite{DL98a, DL98b}).
However, although the spinning dut emission is in good agreement 
with the {\it WMAP} emission at 7.3 and 9.1 mm, it 
is inconsistent with the 3.2 mm emission.
\item The VSGs long-wavelength emission.
VSGs are transiently heated when
an ultraviolet photon is absorbed. The mean interval between
successive ultraviolet photons is longer than the cooling time
and thus, between 2 impacts, the temperature of the particles 
is very low (but is at least the CMB temperature). Such particles
could therefore emit significant emission in the millimeter
channels.
\end{itemize}
The models have large uncertainties linked to the unknown properties
of the small particles. It is therefore very difficult to predict
the exact contribution of the two in the millimeter.

\begin{acknowledgements}
The author thanks the {\it WMAP} team for having provided
to the community beautiful data. Many thanks to J.-L. Puget
and F. Boulanger for having carefully read this paper
and for fruitful discussions.
Thanks also to J.-P. Bernard for his help
in the data manipulation.\\
\end{acknowledgements}


\begin{thebibliography}{}
\bibitem[1994]{Abergel94} Abergel A., Boulanger F., Mizuno A. \& Fukui Y., 1994, ApJ 423, L59 
\bibitem[1991]{Banday91} Banday A.J.\& Wolfendale A.W., 1991, MNRAS 248, 705
\bibitem[2003]{Banday03} Banday A.J., Dickinson C., Davies R.D. et al., 2003, MNRAS submitted
\bibitem[1993]{Bernard93} Bernard J.P., Boulanger F. \& Puget J.-L., 1993, ApJ 277, 609
\bibitem[1992]{Bennett92} Bennett C.L., Smoot G.F. \& Hinshaw G., 1992, 396, L7
\bibitem[2003]{Bennett03} Bennett C.L., Hill R.S., Hinshaw G. et al., 2003, ApJ, submitted
\bibitem[1996]{Boulanger96} Boulanger F., Abergel A., Bernard J.-P. et al. 1996, A\&A 312, 256 
\bibitem[1990]{Boulanger90} Boulanger F., Falgarone E., Puget J.-L. \& Helou G., 1990, ApJ 364, 136
\bibitem[1990]{Desert90} D\'esert F.-X., Boulanger F., Puget J.-L., 1990, A\&A 327, 215 
\bibitem[1999]{DOC99} De Oliveira-Costa A., Tegmark M., Gutierrez C. et al., 1999, ApJ 527, L9 
\bibitem[2002]{DOC02} De Oliveira-Costa A., Tegmark M., Finkbeiner D., et al., 2002, ApJ 567, 363
\bibitem[2003]{Dickinson03} Dickinson C., Davies R.D. \& Davis R.J., 2003, MNRAS, in press
\bibitem[1985]{Draine85} Draine B.T. \& Anderson N. 1985, ApJ 292, 494 
\bibitem[1998a]{DL98a} Draine B.T. \& Lazarian A., 1998a, ApJ 494, L19
\bibitem[1998b]{DL98b} Draine B.T. \& Lazarian A., 1998b, ApJ 508, 157
\bibitem[1999]{DL99} Draine B.T. \& Lazarian A., 1999, ApJ 512, 740
\bibitem[1997]{Dwek97} Dwek E., Arendt, R.G., Fixsen, D.J. et al. 1997, ApJ 475, 565 
\bibitem[1999]{Finkbeiner99} Finkbeiner D., Davis M. \& Schlegel D.J, 1999, ApJ 524, 867
\bibitem[2002]{Finkbeiner02} Finkbeiner D., Schlegel D.J, Frank C.\& Heiles C., 2002, ApJ 566, 898
\bibitem[2003]{Finkbeiner03} Finkbeiner D., 2003, ApJS, in press
\bibitem[2001]{Gaustad01} Gaustad J.E., Mc Cullough P.R., Rosing W.R. \& Buren D.V., 2001, PASP 113, 1326
\bibitem[1999]{Haffner99} Haffner L.M., 1999, Ph.D. Thesis, University of Wisconsin
\bibitem[1994]{Hartmann94} Hartmann D., 1994, Ph.D. Thesis, University of Leiden
\bibitem[1997]{Hartmann97} Hartmann D.\& Burton W.B., Cambridge University Press, 1997
\bibitem[1982]{Haslam82} Haslam C.G.T., Salter C.J., Stoffel H. \& Wilson W., 1982, A\&AS 47, 1
\bibitem[1996]{Kogut96} Kogut A., Banday A.J., Bennett C.L. et al., 1996, ApJ 460, 1 
\bibitem[1999]{Kogut99} Kogut A., 1999, in ASP Conf. Ser. 181, Microwave Foregrounds, ed. 
A. De Oliveira Costa \& M. Tegmark, in press
\bibitem[1998]{Lagache98} Lagache G., Abergel A., Boulanger F. \& Puget J.-L., 1998, A\&A 333, 709
\bibitem[1999]{Lagache99} Lagache G., Abergel A., Boulanger F., et al., 1999, A\&A 344, 322
\bibitem[2003]{Lagache03} Lagache G., 2003, ``Absolute photometric calibration of {\it Planck/HFI} high-frequency 
channels on the Galaxy: Application to Archeops data'', A\&A, to be sumitted
\bibitem[1991]{Laureijs91} Laurejis R.J., Clark F.O. \& Prusti T., 1991, ApJ 372, 185 
\bibitem[2001]{Li01} Li A. \& Draine, B.T., 2001, ApJ 554, 778 
\bibitem[1995]{Puget85} Puget, J.L., L\'eger, A. \& Boulanger, F. 1985, A\&A 142, L19
\bibitem[1995]{Reach95} Reach W.T., Dwek E., Fixsen D.J., et al. 1995, ApJ 451, 188
\bibitem[1998]{Reynolds98} Reynolds R.J., Tufte S.L., Haffner L.M., et al., 1998, PASA 15, 14
\bibitem[1992]{Sieben92} Siebenmorgen R. \& Kr\"ugel, E. 1992, A\&A 259, 614
\bibitem[2003]{Stepnik03} Stepnik B., Abergel A., Bernard J.-P.et al., 2003, A\&A 398, 551
\bibitem[1986]{Weiland86} Weiland J.L., Blitz L., Dwek E. et al. 1986, ApJ 306, L101 

\end{thebibliography}
\end{document}